
\input phyzzx
\date{\vbox{\hbox{November 1992}\par\hbox{KANAZAWA-92-19}}}
\titlepage
\hoffset=0mm\hsize=160mm
\voffset=3mm\vsize=240mm

\title{ Two Dimensional Black Hole Evapolation in the Light-Cone Gauge
\footnote*{Reseach supported in part by SASAKAWA SCIENTIFIC RESEARCH
GRANT of THE JAPAN SCIENCE SOCIETY}
}
\author{Haruhiko Terao
\footnote\dagger{e-mail:terao@hep.s.kanazawa-u.ac.jp}
}
\vskip 0.5cm
\address{Department of Physics, Kanazawa University \break Kanazawa 920-11,
 Japan}

\abstract{
 Quantization of the pure $1+1$ dimensional dilaton gravity is examined
in the light-cone gauge.
It is found that the total action including ghosts generates a $c=0$ free
 conformal field theory
without modification of the classical action, which is required
in the conformal gauge.
We also study semiclassical equations of the dilaton gravity coupled
to $N$ scalar fields.
It is shown that the black hole singularity is not removed even
for $N<24$ in the light-cone gauge.
This indicates that the semiclassical analysis breaks down for small $N$.
}

\endpage

The interest of studying two dimensional black hole physics has been
growing recently. A paticular interesting toy model describing the
black hole formation and evapolation was proposed by Callan,
Giddings, Harvey and Strominger.
\Ref\CGHS{C. G. Callan, S. B. Giddings, J. A. Harvey and
A. Strominger, Phys. Rev. D45 (1992) R1005.}
The system is two dimensional gravity coupled to a dilaton $\phi$
and $N$ scalar fields $\varphi_i$ described by the classical action
$$
S^{class.}={1 \over 2\pi}\int d^2x \sqrt{-g} \left[ e^{-2\phi}
\left( R+4(\nabla\phi)^2 + 4\lambda^2
\right) -{1\over2}\sum_{i=1}^N (\nabla\varphi_i)^2 \right],
\eqno(1)
$$
which resembles the target space action of $c=1$ non-critical string
theory. It was conjectured that the collapsing matter radiates away
all of its energy before an event horizon is formed and black holes
thereby disappear from the quantum mechanical spectrum. Subsequently
it was shown that the gravitational collapse always develops a
singularity for large $N$ and the semiclassical equations must break
down before the singularity is reached.
\Ref\BDDO{T. Banks, A. Dabholkar, M. R. Douglas and M. O'Loughlin,
Phys. Rev. D45 (1992) 3607.}
\Ref\RST{J. G. Russo, L. Susskind and L. Thorlacius, Phys. Lett. B292
(1992) 13.}

However Strominger
\Ref\S{A. Strominger, Phys. Rev. D46 (1992) 4396.}
proposed a modification of the dilaton gravity action so that dilaton,
graviton or ghost does not contribute to the Hawking radiation.
In order to avoid such Hawking radiation it is natural to define the
measure of graviton-dilaton-ghost system with respect to the rescaled
metric
$\hat{g}_{\mu\nu} =e^{-2\phi}g_{\mu\nu}$.
It should be noted that the measure of the matter fields must be
defined by using the physical metric $g_{\mu\nu}$, otherwise we can
not obtain Hawking radiation at all. The semiclasscal action turns out
to be in the conformal gauge $g_{\mu\nu} = e^{2\rho}\eta_{\mu\nu}$ as
$$
\eqalign{
S={1\over\pi}\int d^2x \bigg[&
e^{-2\phi}\left( 2\partial_+\partial_-\rho - 4\partial_
+\phi\partial_-\phi + \lambda^2 e^{2\rho} \right) \cr
&-{N\over 12}\partial_+\rho\partial_-\rho + {1\over2}\sum_{i=1}^N
\partial_+\varphi_i\partial_-\varphi_i
+ 2\partial_+(\rho-\phi)\partial_-(\rho-\phi)  \bigg].
\cr}
\eqno(2)
$$
It was also pointed out that the dilaton-gravity sector without the
cosmological constant term gives a free $c=26$ conformal field
theory in terms of the variables
$\eta=e^{-2\phi}$ and $\hat{\rho}=\rho-\phi$
due to the presence of the last induced term. In the ref. [4] it
was shown that the modification accounting the measures removes the
black hole singularities if $N<24$ in the semiclassical approximation.

In this letter such semicalssical analysis of the dilaton gravity
coupled $N$ scalar fields defined by the action (1) is examined in
the light-cone gauge. We  shall find the modification performed in
the ref. [4] is not necessary in order to avoid the Hawking radiation
of gravity-dilaton-ghost modes. Consequently it will be also shown
that the singularity still appears even for $N<24$.

First we shall examine quantization of the pure gravity coupled to
dilaton. The measure should be defined with respect to the rescaled
metric $\hat{g}_{\mu\nu}$. Therefore it would make analysis to be
transparent to rewrite the action (1) in terms of the rescaled metric.
Then the pure dilaton-gravity part of (1) turns out to be
$$
S={1\over2\pi}\int d^2x \sqrt{-\hat{g}}\left[ e^{-2\phi}\hat{R}
+ 4\lambda^2 \right] .
\eqno(3)
$$
Then dilaton acts as a Lagrange multiplier which enforces the rescaled
metric to be always flat; $\hat{R}=0$. Actually the action (3) is very
similar to the gravitational model proposed by Jackiw and Teitelboim.
\Ref\JT{R. Jackiw, in: Quantum Theory of Gravity, ed. S. Christensen
(Adam Hilger Bristol, 1984) p.403,
\nextline C. Teitelboim, in: Quantum Theory of Gravity,
ed. S. Christensen (Adam Hilger Bristol, 1984) p.327.}

Here we impose the light-cone gauge
\Ref\KPZ{A. M. Polyakov, Mod. Phys. Lett. A2 (1987) 893,
\nextline V. G. Knizhnik, A. M. Polyakov and A. B. Zamolodchikov,
Mod. Phys. Lett. A3 (1988) 819.}
to the rescaled metric as
$$
\hat{g}_{--} = 0, \qquad\hat{g}_{+-}=\hat{g}_{-+} = -{1\over2},
\eqno(4)
$$
where the definition of the light-cone coordinates follows the notation
in the ref.[1]. Quantization of Jackiw and Teitelboim's model in the
light-cone gauge has been studied by the author.
\Ref\T{H. Terao, preprint DPKU-9207, hep-th/9205030.}
It will be seen that we may proceed quantization of the dilaton-gravity
in a very similar way. Introducing Fadeev-Popov ghosts $c^{\pm}$ and
antighosts $b_{++}$ and $b_{+-}$, the gauge fixed action will be given
by
$$
S ={1\over\pi}\int dx^{\pm} \bigg\{
\eta\partial_-^2\hat{g}_{++} + \lambda^2 + b_{++}\partial_-c^+ +
 b_{+-}( \partial_+c^+ + \partial_-c^-) \bigg\},
\eqno(5)
$$
where $\eta$ denotes $e^{-2\phi}$. If we regard $\eta$ as an elementary
field, then the action (5) is free including the cosmological constant
term. Indeed the correct measure of the dilaton should be determined
by consistency. Therefore we shall treat $\eta$ as a free field and
will verify the consistency later. One more remark should be mentioned.
As $\phi$ ranges $-\infty$ to $\infty$, $\eta$ ranges from $0$ to $\infty$.
To proceed quantization we will assume that such boundary condition is
harmless for the consistency.

The action (5) is found to be invariant under the following BRST
transformations,
$$
\eqalign{
&\delta^B \hat{g}_{++}
=-\partial_+c^- + c^+\partial_+\hat{g}_{++}
+ c^-\partial_-\hat{g}_{++} + 2\partial_+c^+\hat{g}_{++}, \cr
&\delta^B \eta
=c^+\partial_+\eta + c^-\partial_-\eta,\cr
&\delta^B c^{\pm}
=c^+\partial_+c^{\pm} + c^-\partial_-c^{\pm},\cr
&\delta^B b_{++}
=T'^{DG}_{++} + T'^{Gh}_{++} \equiv T'^{tot}_{++}, \cr
&\delta^B b_{+-}
=T'^{DG}_{+-} + T'^{Gh}_{+-} \equiv T'^{tot}_{+-}, \cr
}\eqno(6)
$$
where $T'_{++}$ and $T'_{+-}$ are given explicitly by
$$
\eqalign{
T'^{DG}_{++}&=
- \left( \partial_+ - \partial_-\hat{g}_{++}
+ 2\hat{g}_{++}\partial_- \right)\partial_+\eta
- \partial_+\hat{g}_{++}\partial_-\eta,\cr
T'^{DG}_{+-}&=
\left(\partial_+ + \partial_-\hat{g}_{++}\right)
\partial_-\eta +\lambda^2,\cr
T'^{Gh}_{++}&=
c^+\partial_+b_{++}  + 2 \partial_+c^+b_{++}
+ c^-\partial_-b_{++}, \cr
T'^{Gh}_{+-}&=
c^+\partial_+b_{+-} + c^-\partial_-b_{+-}. \cr
}\eqno(7)
$$
$T'^{DG}_{++}$ and $T'^{DG}_{+-}$ in (7) are related to the energy
-momentum tensor of the dilaton-gravity system given by
$$
T_{\mu\nu}=-{2\pi \over \sqrt{-\hat{g}}}
{\delta S \over \delta \hat{g}^{\mu\nu}}.
\eqno(8)
$$
Indeed the energy-momentum tensor is found to be \refmark{\T}
$$
\eqalign{
T^{DG}_{++}&=
(-2\hat{g}_{++})^2 T^{DG}_{--} + 2\hat{g}_{++}T'^{DG}_{+-}
- T'^{DG}_{++},\cr
T^{DG}_{+-}&
=-2\hat{g}_{++}T^{DG}_{--} - T'^{DG}_{+-}, \cr
T^{DG}_{--}&=\partial_-^2 \eta.	\cr
}\eqno(9)
$$
It is also straightforward to derive the conserved BRST currents
corresponding to the BRST transformations (6). They may be written
down in terms of $T'_{++}$ and $T'_{+-}$ as
$$
\eqalign{
J_{+}^{BRST}&=
c^+\left( T'^{DG}_{++} + {1\over2}T'^{Gh}_{++} \right) +
c^-\left( T'^{DG}_{+-} + {1\over2}T'^{Gh}_{+-}\right), \cr
J_{-}^{BRST}&=
c^+ T'^{DG}_{+-}.\cr
}\eqno(10)
$$

Now we may proceed quantization explicitly, since the action (5)
is free. The solutions of the equations of motion from (5) may be
given by
$$
\eqalign{
&\hat{g}_{++} =
h_{++}(x^+) + x^-h_+(x^+),\cr
&\eta         =
\omega(x^+) + x^-\omega^+(x^+), \cr
&c^+          =
\hat{c}^+(x^+), \cr
&c^-          =
\hat{c}^-(x^+) - x^-\partial_+ \hat{c}^+(x^+),\cr
&b_{++}       =
\hat{b}_{++}(x^+) - x^-\partial_+\hat{b}_{+-}(x^+), \cr
&b_{+-}       =
\hat{b}_{+-}(x^+). \cr
}\eqno(12)
$$
Canonical commutation relations lead us to operator product
expansions between the holomorphic fields introduced in (12),
$$
\eqalign{
&\omega^+(x^+) h_{++}(y^+)       \sim {1 \over x^+ - y^+}, \cr
&h_+ (x^+) \omega(y^+)           \sim {1 \over x^+ - y^+}, \cr
&\hat{c}^+(x^+)\hat{b}_{++}(y^+) \sim {1 \over x^+ - y^+}, \cr
&\hat{c}^-(x^+)\hat{b}_{+-}(y^+) \sim {1 \over x^+ - y^+}. \cr
}\eqno(13)
$$
Also if we substitute (12) into (7), then the holomorphic energy
-momentum tensor may be obtained as
$$
\eqalign{
T'^{DG(Gh)}_{++} &=
\hat{T}^{DG(Gh)}_{++}(x^+) - x^-\partial_+
\hat{T}^{DG(Gh)}_{+-}(x^+),  \cr
T'^{DG(Gh)}_{+-} &=
\hat{T}^{DG(Gh)}_{+-}(x^+),  \cr
}\eqno(14)
$$
where
$$
\eqalign{
\hat{T}^{DG}_{++}(x^+) &=
-2h_{++}\partial_+\omega^+ - \partial_+h_{++}\omega^+
-\partial^2_+\omega + h_+\partial_+\omega, \cr
\hat{T}^{DG}_{+-}(x^+) &=
\partial_+\omega^+ + h_+\omega^+ + \lambda^2, \cr
\hat{T}^{Gh}_{++}(x^+) &=
2\partial_+\hat{c}^+\hat{b}_{++}
+ \hat{c}^+\partial_+\hat{b}_{++}
-\hat{c}^-\partial_+\hat{b}_{+-},  \cr
\hat{T}^{Gh}_{+-}(x^+) &=
\hat{c}^+\partial_+\hat{b}_{+-}.   \cr
}\eqno(15)
$$
These are indeed the conserved currents for the residual symmetry
after imposing the light-cone gauge. \refmark{\T}

Now the quantum algebras between these currents can be derived by
using the O.P.E.'s given by (13). After some manipulations they
are found to be
$$
\eqalign{
\hat{T}^{DG}_{++}(x^+)\hat{T}^{DG}_{++}(y^+) &\sim
  {28/2 \over (x^+ - y^+)^4}
+ {2\hat{T}^{DG}_{++}(y^+) \over (x^+ - y^+)^2} +
  {\partial_+\hat{T}^{DG}_{++}(y^+) \over x^+ - y^+}, \cr
\hat{T}^{DG}_{+-}(x^+)\hat{T}^{DG}_{+-}(y^+) &\sim 0, \cr
\hat{T}^{DG}_{++}(x^+)\hat{T}^{DG}_{+-}(y^+) &\sim
  {\partial_+\hat{T}^{DG}_{+-}(y^+) \over x^+ - y^+}  \cr
}\eqno(16)
$$
and
$$
\eqalign{
\hat{T}^{Gh}_{++}(x^+)\hat{T}^{Gh}_{++}(y^+) &\sim
  {-28/2 \over (x^+ - y^+)^4}
+ {2\hat{T}^{Gh}_{++}(y^+) \over (x^+ - y^+)^2} +
  {\partial_+\hat{T}^{Gh}_{++}(y^+) \over x^+ - y^+}, \cr
\hat{T}^{Gh}_{+-}(x^+)\hat{T}^{Gh}_{+-}(y^+) &\sim 0, \cr
\hat{T}^{Gh}_{++}(x^+)\hat{T}^{Gh}_{+-}(y^+) &\sim
  {\partial_+\hat{T}^{Gh}_{+-}(y^+) \over x^+ - y^+}. \cr
}\eqno(17)
$$
Thus it is seen that the central charge of the Virasoro algebras
for the dilaton-gravity sector and the ghost sector are found to
be $28$ and $-28$ respectively. Therefore this dilaton-gravity system
suffers no conformal anomaly without any modification.

Recently it has been pointed out by Cangemi and Jackiw
\Ref\CJ{D. Cangemi and R. Jackiw, Phys. Rev. Lett. 69 (1992) 233.}
that the dilaton-gravity action (3) may be reformulated as a
topological gauge theory based on the Poincar\'e group with central
extension. The absence of anomaly would be expected naively from this
observation. Also it should be reminded that the modification is
required to achieve anomaly cancelation in the conformal gauge.
This difference seems to come from different choice of the
gravitational variable. Of course it would not be expected for
these two formulations of quantum dilaton-gravity to give rise to
different physics. However it will be seen that the argument of the
black hole evapolation based on the semiclassical analysis must be
affected.

We are now in a position to discuss two dimensional black hole
formation and evapolation by incorporating $N$ scalar fields.
The classical equations of motion from the action (1) are easily found
to give same black hole solutions as those obtained in the conformal
gauge. The action (1) turns out to be in the light-cone gauge
$$
S^{class.}={1\over\pi}\int dx^{\pm}
\bigg\{
\eta\partial_-^2\hat{g}_{++} + \lambda^2 +
{1\over2}\sum_{i=1}^N  (\partial_+
+ \hat{g}_{++}\partial_-)\varphi_i\partial_-\varphi_i
\bigg\}.
\eqno(18)
$$
Variation by the fields $\hat{g}_{++}$, $\eta$ and $\varphi_i$
lead the following equations of motion,
$$
\eqalign{
&\partial_-^2 \eta +{1\over2}\sum_{i=1}^N
\partial_-\varphi_i\partial_-\varphi_i = 0, \cr
&\partial_-^2\hat{g}_{++}=0, \cr
&\partial_+\partial_-\varphi_i
+ \partial_-(\hat{g}_{++}\partial_-\varphi_i) = 0. \cr
}\eqno(19)
$$
We must also impose the energy-momentum tensor as constraints,
which come from the equations of motion of $\hat{g}_{+-}$ and
$\hat{g}_{--}$. They are given by
$$
\eqalign{
&(\partial_+ + \partial_-\hat{g}_{++})\partial_-\eta
+ \lambda^2 = 0, \cr
&(\partial_+ - \partial_-\hat{g}_{++} +
2\hat{g}_{++}\partial_-)\partial_+\eta +
\partial_+\hat{g}_{++}\partial_-\eta +
{1\over2}\sum_{i=1}^N  (\partial_+ +
2\hat{g}_{++}\partial_-)\varphi_i\partial_+\varphi_i =0.\cr
}\eqno(20)
$$
The second equation of (19) implies that $\hat{g}_{++}$ gives a flat
metric. Therefore it is always possible to make $\hat{g}_{++}$ to
vanish by coordinate transformation. Then the other equations reduce
to
$$
\eqalign{
&\partial_+^2 \eta =
-{1\over2}\sum_{i=1}^N \partial_+\varphi_i\partial_+\varphi_i ,\cr
&\partial_-^2 \eta =
 -{1\over2}\sum_{i=1}^N \partial_-\varphi_i\partial_-\varphi_i , \cr
&\partial_+\partial_-\eta = -\lambda^2, \cr
&\partial_+\partial_-\varphi_i  = 0, \cr
}\eqno(21)
$$
which are identical to the equations obtained in the conformal gauge.
\refmark\CGHS If we take the matter to be a shock-wave traveling in
the $x^-$ direction described by
${1\over2}\sum_{i=1}^N \partial_+\varphi_i\partial_+\varphi_i$
$= a\delta(x^+ - x^+_0)$,
then the solution of the equations of (21) is given by
$$
\eta = e^{-2\phi}
= -a(x^+ - x^+_0)\Theta(x^+ - x^+_0) - \lambda^2x^+x^-.
\eqno(22)
$$
Since the space-time geometry is described by the physical metric
$g_{\mu\nu} = e^{2\phi}\hat{g}_{\mu\nu}$,
this solution indeed gives a black hole geometry.\refmark\CGHS

In order to discuss the Hawking radiation and the black hole
evapolation we have to include the quantum corrections by the
matter fields. At one-loop level the matter fields contribute the
well-known conformal anomaly term
$$
S^{corr.} = -{N\over96\pi}\int d^2x \sqrt{-g} R {1\over\Delta} R
\eqno(23)
$$
to the effective action.
\Ref\P{A. M. Polyakov, Phys. Lett. B163 (1981) 207.}
 Here it should be noted that this non-local action is given in
terms of the physical metric $g_{\mu\nu}$, not by the rescaled
metric $\hat{g}_{\mu\nu}$, because the measures of the matter are
defined by using $g_{\mu\nu}$.\refmark\S If we rewrite (23) in
terms of  $\hat{g}_{\mu\nu}$ and $\phi$, then we obtain
$$
S^{corr.} = -{N\over96\pi}\int d^2x
\left\{
\sqrt{-\hat{g}} \hat{R} {1\over\hat{\Delta}} \hat{R} -
4\sqrt{-\hat{g}} \hat{R}\phi -4\sqrt{-\hat{g}} (\hat{\nabla}\phi)^2
\right\}.
\eqno(24)
$$
Now we shall examine the semiclassical equations of motion derived
from the effective action
$S^{eff.} = S^{class.} + S^{corr.}$
in the light-cone gauge (4). Introducing a new variable $f$ by
$\hat{g}_{++} \equiv -\partial_+f / \partial_-f$,
the anomaly term (24) may be rewritten into
$$
S^{corr.} = -{N\over6}\Gamma_+[f] +
{N\over12\pi}\int dx^{\pm}
\left\{
\phi\partial^2_-\hat{g}_{++} -
(\partial_+ + \hat{g}_{++}\partial_-)\phi\partial_-\phi
\right\}.
\eqno(25)
$$
Here $\Gamma_+[f]$ is so-called the gravitational WZNW action
\refmark{\KPZ}
\Ref\CR{A. H. Chamseddine and M. Reuter, Nucl. Phys. B317 (1989) 757.}
which is given by
$$
\Gamma_+[f] = {1\over8\pi} \int dx^{\pm}
\left[
{\partial^2_-f \partial_+\partial_-f \over (\partial_-f)^2} -
{\partial_+f \over \partial_-f}
\left({\partial_-^2f \over \partial_-f }\right)^2
\right].
\eqno(26)
$$

The flux of Hawking radiation from a black hole formed by collapse
of the infalling shock-wave may be obtained by evaluating the
energy-momentum tensor for the anomaly term with respect to the
black hole background (22). Through a bit tedious but staightforward
calculations we may find the energy-momentum tensor to be
$$
\eqalign{
T^P_{--} = &
-{N\over24}\left[ {\partial_-^3f \over \partial_-f}
- {3\over2}\left( {\partial_-^2f \over \partial_-f}
\right)^2 \right]
+{N\over12}\left[ \partial_-^2\phi
- \partial_-\phi\partial_-\phi \right], \cr
T^P_{+-} = &
-\hat{g}_{++}T^P_{--} - {N\over24}\partial_-^2\hat{g}_{++}
+{N\over12}\left[ -\hat{g}_{++}\partial_-^2\phi
- (\partial_+ + \partial_-\hat{g}_{++})\partial_-\phi \right], \cr
T^P_{++} = &
\hat{g}_{++}^2T^P_{--}
- {N\over48}\left[ -6\hat{g}_{++}\partial_-^2\hat{g}_{++}
+ (\partial_-\hat{g}_{++})^2
- 2 \partial_+\partial_-\hat{g}_{++} \right] \cr
&+{N\over12} [
3\hat{g}_{++}^2\partial_-^2\phi
+ 2\hat{g}_{++}(\partial_+
+ \partial_-\hat{g}_{++})\partial_-\phi
+ \partial_+\hat{g}_{++}\partial_-\phi \cr
&\qquad + (\partial_+ - \partial_-\hat{g}_{++}
+ 2\hat{g}_{++}\partial_-)\partial_+\phi
-(\partial_+\phi + 2 \hat{g}_{++}\partial_-\phi)\partial_+\phi
].\cr
}\eqno(27)
$$
However after inserting the classical black hole solution
$\hat{g}_{++} =0$ (or $f=0$), they reduce to
$$
\eqalign{
T^P_{--} = &
{N\over12}\left[ \partial_-^2\phi - (\partial_-\phi)^2 \right],\cr
T^P_{+-} = &
-{N\over12}\partial_+\partial_-\phi , \cr
T^P_{++} = &
{N\over12}\left[ \partial_+^2\phi - (\partial_+\phi)^2 \right].\cr
}\eqno(28)
$$
Therefore it is now apparent to reproduce Hawking radiation calculated
in the conformal gauge. \refmark{\CGHS}

Lastly we shall examine the semiclassical equations of motion
derived from the effective action $S^{eff.}$, which are sometimes
refered as the CGHS equations in the conformal gauge. Especially
we are interested in the curvature singularity appearing in the
CGHS equations. Therefore we would like to focus our attention on
the following equations among several ones,
$$
\eqalign{
&2\left( e^{-2\phi} - {N\over24} \right)\partial_-^2\hat{g}_{++}
= {N\over6} \left[\partial_+\partial_-\phi
- \partial_-(\hat{g}_{++}\partial_-\phi) \right],  \cr
&(\partial_+ + \partial_-\hat{g}_{++} + \hat{g}_{++}\partial_-)
\partial_- \left( e^{-2\phi} + {N\over12}\phi \right) + \lambda^2
= -{N\over24}\partial_-^2\hat{g}_{++}. \cr
}\eqno(29)
$$
On the other hand the physical curvature $R$ is rewritten into
$$
R = 4e^{-2\phi} \left[
\partial_-^2\hat{g}_{++}
+ 2\partial_-(\partial_+\phi +\hat{g}_{++}\partial_-\phi)
\right]
\eqno(30)
$$
in terms of $\hat{g}_{++}$ and $\phi$. After eliminating the second
derivatives in the expression (30) by using the equations given in
(28), we may easily obtain
$$
R = {1\over 1-{N\over12}e^{2\phi}}
\left[
4e^{-2\phi}(\partial_+\phi
+ \hat{g}_{++}\partial_-\phi) \partial_-\phi
+ \lambda^2 \right].
\eqno(31)
$$
Thus the curvature singularity appears certainly when the dilaton
approaches to $\phi={1\over2}ln{12 \over N}$. This holds even for
$N<24$, which should be compared with the results in the conformal
gauge.\refmark{\S} It is somehow curious that existence of the quantum
black hole singularity depends on the gauge choice. Rather this result
seems to indicate that such semiclassical analysis can not be justified
for small $N$. Therefore full quantum treatment of the dilaton gravity
coupled to matter would be necessary to analyze the black hole
evapolation. Some attempts have been performed.
\Ref\QBH{A. Bilal and C. Callan, preprint PUPT-1320, hep-th/9205089,
\nextline S. P. de Alwis, preprint COLO-HEP-284, hep-th/9206020
and COLO-HEP-288, hep-th/9207095,
\nextline K. Hamada, preprint UT-Komaba 92-7, hep-th/9206071.}
However the physics of quantum black holes has not been completely
clear yet and the analysis in the light-cone gauge may be useful
in this direction also.

Note added; Recently the model for dilaton gravity (1) has been
examined in the light-cone gauge also by Shen, while the gauge
conditions are impodsed on the physical metric.
\Ref\SH{X. Shen, preprint CERN-TH 6633/92, hep-th/9209064.}

\ack
The author is grateful to Prof. J. Kubo and Dr. I. Oda for valuable
discussions.

\refout
\bye